\documentclass[aps,prl,twocolumn,showpacs,preprintnumbers,superscriptaddress,amsmath,amssymb]{revtex4}
\usepackage{bm,mathrsfs,eucal,graphicx}
\usepackage[hypertex]{hyperref}
\renewcommand{\vec}[1]{\bm{#1}}
\begin{document}

\title{Quantum effects for the 2D soliton in isotropic ferromagnets}

\author{Boris A. Ivanov} %
\affiliation{Institute of Magnetism, NASU, 03142 Kiev, Ukraine} %
\affiliation{National Taras Shevchenko University of Kiev, 03127 Kiev,
Ukraine}

\author{Denis D. Sheka}
\email{Denis\_Sheka@univ.kiev.ua} %
\affiliation{National Taras Shevchenko University of Kiev, 03127 Kiev,
Ukraine}

\author{Vasilii V. Krivonos}
\affiliation{National Taras Shevchenko University of Kiev, 03127 Kiev,
Ukraine}

\author{Franz G. Mertens}
\affiliation{Physikalisches Institut, Universit\"at Bayreuth, D--95440
Bayreuth, Germany}

\date{January 22, 2007}
%\preprint{{\texttt{version 3}}}

\begin{abstract}

We evaluate a zero--point quantum correction to a Belavin--Polyakov soliton in
an isotropic 2D ferromagnet. By revising the scattering problem of
quasi--particles by a soliton we show that it leads to the Aharonov--Bohm type
of scattering, hence the scattering data can not be obtained by the Born
approximation. We proof that the soliton energy with account of quantum
corrections does not have a minimum as a function of its radius, which is
usually interpreted as a soliton instability. On the other hand, we show that
long lifetime solitons can exist in ferromagnets due to an additional integral
of motion, which is absent for the $\sigma$--model.
\end{abstract}

%
% 75.10.Hk -  Classical spin models
% 75.30.Ds -  Spin waves
% 05.45.Yv -  Solitons
%

\maketitle

Solitons are known to play an important role in several branches of field
theory and condensed matter physics, see Ref.~\onlinecite{Manton04} for a
review. In particular, solitons treated as nonlinear excitations are important
in 1D and 2D magnetism \cite{Kosevich90,Mikeska91,Baryakhtar93}. A serious
impediment in studying 2D spin systems arises due to the absence of exact
analytical solutions for most models. Thereupon special attention is deserved
to models which admit an analytical treatment. One of the well--known examples
is a model of the 2D isotropic ferromagnet (FM), which provides an exact
analytical soliton, the so--called Belavin--Polyakov (BP) soliton
\cite{Belavin75}. In terms of the normalized magnetization, $\vec{m} = \left(
\sin\theta\cos\phi, \sin\theta\sin\phi, \cos\theta\right)$, the soliton
structure $\vec{m}_{\text{BP}}$ is described by the formula \cite{Belavin75}
\begin{equation} \label{eq:BP-soliton}
\tan \frac{\theta_{\text{BP}}}{2} = \left(\frac{R}{\rho}\right)^{|q|}, \qquad
\phi_{\text{BP}} = \varphi_0 + q\chi.
\end{equation}
Here $\rho$ and $\chi$ are the polar coordinates in the magnet plane, the
integer $q$ is the $\pi_2$--topological charge of the soliton, $R$ and
$\varphi_0$ are arbitrary parameters. BP--solitons of the form
\eqref{eq:BP-soliton} appear in different models of non-linear  field theory
and condensed matter physics \cite{Manton04}. In particular, the BP--solitons
are important for the ferromagnetic quantum Hall effect \cite{Abolfath98}.

The unique problem of the BP soliton is that its energy
$\mathscr{E}_{\text{BP}}=4\pi JS^2|q|$ ($J$ is the exchange integral, $S$ is
the atomic spin) does not depend on its radius $R$: this results from the
scale invariance of the system, which is a part of the general conformal
invariance of the model. Since the soliton radius is not fixed, one is then
free to let it go up to the system size, and the thermal excitation of
solitons will break the long-range order \cite{Belavin75}. However, recent
studies have shown that a quantization of the soliton of the classical
$\sigma$-model, which can be attributed to antiferromagnets, breaks the static
scale invariance \cite{Rodriguez89,Moussallam91,Walliser00}. The natural
question whether it works for FMs is still open.

The purpose of our study is to examine the role of quantum fluctuations for
the soliton properties. We treat the problem semiclassically using the
one--loop correction to the classical soliton energy, originally calculated by
\citet{Dashen74I,Dashen74II} for 1D solitons, see also
Ref.~\onlinecite{Rajaraman82}. To generalize these results to the 2D case one
needs to solve the soliton--magnon scattering problem for 2D magnets. We show
a \emph{unique new feature} of the 2D soliton--magnon interaction, which is
absent in 1D: the soliton acts on magnons not only by some local potential,
but also in the same way as an effective long--ranged magnetic field acts on a
charged particle; this essentially changes the scattering picture, leading to
the Aharonov--Bohm (AB) scenario. We state that the AB type of scattering is a
general consequence of 2D scattering by a topological soliton. We calculate
the Casimir energy for the BP--soliton in FMs and show that the quantum
correction can not provide a fixed size for the soliton. Nevertheless, we show
that long lifetime solitons can exist in FMs due to an additional integral of
motion, contrary to antiferromagnets.

The macroscopic dynamics of the classical FM follows the Landau--Lifshitz
equations
\begin{equation} \label{eq:LL}
\begin{split}
\frac{1}{D}\sin\theta\ \partial_t \phi &= \vec{\nabla}^2\theta -
\sin\theta\cos\theta\left(\vec\nabla\phi \right)^2,\\
\frac{1}{D}\sin\theta\ \partial_t \theta &=  - \vec{\nabla} \cdot
\left(\sin^2\theta \vec{\nabla} \phi \right),
\end{split}
\end{equation}
where $D$ is the stiffness coefficient of the spin--waves, which are
characterized by the dispersion law $\omega(\vec{k}) = Dk^2$. To analyze the
soliton--magnon interaction, we consider small oscillations of the
magnetization $\vec{m}$ on the background of the stationary BP--soliton
$\vec{m}_{\text{BP}}$. These oscillations can be described in terms of the
complex valued ``wave function'' $\Psi = \theta-\theta_{\text{BP}} + i
\sin\theta_{\text{BP}} (\phi-\phi_{\text{BP}})$, see
Ref.~\onlinecite{Ivanov99}. For the further analysis it is instructive to
rewrite the linearized equation for the $\Psi$--function in the form of the
Schr\"{o}dinger equation:
\begin{subequations} \label{eq:wave}
\begin{align}
\label{eq:wave-1} %
H\Psi &= \frac{i}{D}\partial_t\Psi, \quad
H = (-i\vec\nabla - \vec{A})^2 + V,\\
\label{eq:wave-2} %
V &= -\frac{q^2}{\rho^2}\sin^2\theta_{\text{BP}}, \qquad \vec{A} =
-\frac{q\cos\theta_{\text{BP}}}{\rho}\vec{e}_\chi.
\end{align}
\end{subequations}

The Hamiltonian $H$ has a form which is typical for  a quantum--mechanical
charged particle in the presence of a scalar potential $V$ and an additional
magnetic field with a vector potential $\vec{A}$. The appearance of an
effective magnetic field is a new feature of the 2D soliton--magnon
interaction, which is always absent in 1D systems. Discerning this effective
magnetic field gives the possibility to draw a number of general conclusions
about soliton--magnon scattering in the 2D case, see below.

For the system \eqref{eq:wave} we apply the standard partial wave expansion,
using the \emph{Ansatz}:
\begin{equation} \label{eq:partial-waves}
\Psi(\rho,\chi,t) = \sum_{\alpha=(k,m)} \psi_m(\rho) \exp\left( im\chi -
i\omega_\alpha t + \beta_\alpha\right).
\end{equation}
Here the integer $m$ is the azimuthal quantum number, $k$ is the radial wave
number, and $\beta_\alpha$ is an arbitrary initial phase. Each partial wave
$\psi_m$ is an eigenfunction of the 2D radial Schr\"{o}dinger equation
\begin{equation} \label{eq:Schroedinger}
\begin{split}
& \left(-\nabla_\rho^2 + U_m \right) \psi_m = k^2 \psi_m,\\
&U_m  = \frac{m^2 + 2mq\cos\theta_{\text{BP}} \! +
q^2\cos2\theta_{\text{BP}}}{\rho^2}.
\end{split}
\end{equation}
Here the term linear in $m$ reminds of an effective magnetic field $\vec{A}$.
The scattering problem can be formulated in the usual way. The eigenfunctions
for free magnon modes have the form $\psi_m^{\text{free}} \propto
J_{|m|}(k\rho) $, with an asymptotic behavior $\psi_m^{\text{free}} \propto
(k\rho)^{-1/2} \cos\left(k\rho - {|m|\pi}/{2} - {\pi}/{4} \right)$ when
$k\rho\gg|m|$; $J_m$ is Bessel function. In the presence of a soliton the
behavior of a magnon solution can be analyzed at large distances, $\rho\gg R$.
In view of the asymptotic behavior $U_m\approx |m+q|^2/\rho^2$, in the
limiting case $k\rho\gg |m|$ one has the usual result \cite{Ivanov99}:
\begin{equation*} \label{eq:delta}
\psi_m \propto \frac{1}{\sqrt{k\rho}} \cos\left( k\rho - \frac{|m+q|\pi}{2} -
\frac{\pi}{4} + \eta_m(k) \right).
\end{equation*}
The phase shift $\eta_m$ contains all information about the scattering
process.

The main features of the scattering on a topological BP--soliton are caused by
the magnetic field $\vec{A}$. As an analogue of the Zeeman splitting of
electron energy terms in an external magnetic field, the presence of an
effective magnetic field breaks the symmetry $\eta_m(k)=\eta_{-m}(k)$. It is
necessary to take into account separately positive and negative $m$'s.

Sometimes the soliton--magnon scattering problem is treated perturbatively
using the Born approximation \cite{Rodriguez89,Walliser00}, which, in
principle, can be used for the scattering in a magnetic field. However, due to
the topological soliton properties, $\vec{A}$ is a long--ranged field
\begin{equation} \label{eq:A-explicit}
\!\!\!\vec{A}(\rho) = \frac{1-\left(\dfrac{\rho}{R}\right)^{2|q|}}{1 +
\left(\dfrac{\rho}{R}\right)^{2|q|}} \frac{q}{\rho}\vec{e}_\chi \sim
  \begin{cases}
    +\dfrac{q}{\rho}\vec{e}_\chi & \text{when $\rho\ll R$}, \\
    -\dfrac{q}{\rho}\vec{e}_\chi & \text{when $\rho\gg R$},
  \end{cases}
\end{equation}
which is typical for the AB effect \cite{Sheka06d}. For such a type of
scattering, some standard scattering results fail. For example, the Levinson
theorem must be modified for long--range potential systems
\cite{Sheka03,Sheka06d}. Since scattering phase shifts are not still
localized, there appears a problem of the regularization of the scattering
series like in conventional AB scattering picture \cite{Henneberger80}. As was
firstly noted by \citet{Feinberg63}, the Born approximation fails for the AB
scattering; it gives an average of two different modes with opposite signs of
$m$ \cite{Ruijsenaars83}. Thus we need a more precise approach than the Born
approximation. One needs to stress that such a long--range behavior is not a
result of slow algebraic decay of the out--of--plane structure of the BP
soliton. It is a consequence of the topology of the BP--soliton, namely, of
the relation $\phi=q\chi$, thus the AB--type of scattering is valid also for
anisotropic magnets \cite{Sheka01,Sheka04,Ivanov05b}.

Let us discuss the soliton with the topological charge $q=1$, which has the
lowest energy. Such a soliton has two internal zero--frequency modes, which
are the limit of the continuum spectrum as $k\to0$: \cite{Ivanov95g}
\begin{equation} \label{eq:psi-zero}
\psi_{m=+1}^{(k=0)} = \frac{1}{\rho^2+R^2}, \qquad \psi_{m=0}^{(k=0)} =
\frac{\rho}{\rho^2+R^2}.
\end{equation}
The mode with $m=+1$ is a local translational mode, which describes a soliton
shift, the mode with $m=0$ is the half--local rotational mode. The mode with
$m=+1$ has an exact analytical solution for any finite values of $k$
\begin{equation} \label{eq:psi4m=+1}
\psi_{m=1}(\rho) = J_2(k\rho) - \frac{2}{k\rho}
\frac{J_1(k\rho)}{(\rho/R)^2+1},
\end{equation}
hence this mode does not scatter at all, $\eta_{m=+1}=0$ \cite{Ivanov99}. Note
that in the interesting case of long--wavelength asymptotic behavior
($k\rho\ll1$) at large distances $\rho \gg R$, this expression has the same
form as a combination of Bessel and Neumann functions, $J_2(k\rho)\propto
(k\rho)^2$ and $Y_2(k\rho)\propto (k\rho)^{-2}$. Thus, the second term in
\eqref{eq:psi4m=+1} imitates the presence of the function $Y_2$ and the
presence of scattering, which caused a conclusion in
Ref.~\onlinecite{Moussallam91} that the mode with $m=+1$ can be scattered. The
scattering phase shift can be found in both limiting cases, for small and
large dimensionless radial wave number $\varkappa=kR$ \cite{Ivanov99}. For
long--wave lengths, $\varkappa \ll 1$
\begin{subequations} \label{eq:delta-asymp}
\begin{equation} \label{eq:delta4k<<1}
\eta_m \underset{\varkappa\ll1}{\sim}
 \begin{cases}
0, & \text{when $m=1$},\\
-\dfrac{\pi}{2\ln(1/\varkappa)}, & \text{when $m=0$},\\
-\pi\varkappa^2\ln\dfrac{1}{\varkappa}, & \text{when $m=-1$},\\
\dfrac{\pi \varkappa^2}{2m(m+1)}\text{sgn }m, & \text{otherwise},
 \end{cases}
\end{equation}
and in the opposite case of short--wave lengths \cite{Ivanov99}
\begin{equation} \label{eq:delta4k>>1}
\eta_m \underset{\varkappa\gg1}{\sim} \pi\,\text{sgn }(m-1) -
\dfrac{\pi(m-1)}{\varkappa}.
\end{equation}
\end{subequations}

Now we are able to calculate the density of magnon states. Let us generalize
the main arguments of \citet{Dashen74I,Dashen74II} for the 2D system. The idea
of the approach is to calculate energy shifts of vacuum magnon states in the
presence of a soliton, which are constructed as one-loop quantum corrections
to the soliton energy. The energy of the vacuum comes from the zero-point
fluctuations of the magnon states. Without the soliton, each vacuum magnon
makes a contribution as $\hslash D\vec{k}_{\text{vac}}^2/2$, where
$\{\vec{k}_{\text{vac}}\}$ is set of allowable wave vectors. In the soliton
presence the set of allowable wave vectors changes, $\{\vec{k}\}$. The energy
of the state with the wave vector $\vec{k}$ is $\hslash D\vec{k}^2/2$.
Therefore the energy correction is
\begin{equation} \label{eq:1-loop}
\mathscr{E}^{\text{1--loop}} = \frac{\hslash D}{2} \sum_{\vec{k}} \vec{k}^2 -
\frac{\hslash D}{2} \sum_{\vec{k}^{\text{vac}}} \vec{k}_{\text{vac}}^2.
\end{equation}
To determine the set of allowable states, we put the system in a very large
box of the size $L$, making all states discrete. In the limiting case
$L\to\infty$ the energy correction \eqref{eq:1-loop} does not depend on the
form of the boundary conditions. For the 2D case we choose fixed boundary
conditions for a circular box of radius $L$ \cite{Ivanov99}. Since the free
magnons are described by the Bessel function $\psi_m^{\text{free}}$, by
enforcing fixed boundary conditions $\psi_m^{\text{free}}(\rho=L)=0$, we fix
the allowed values of the radial wave number, $k_n^{\text{vac}}L=j_m^{(n)}$,
where $j_{m}^{(n)}$ is the $n$--th zero of the Bessel function $J_m$. In the
region of interest, $n\gg 1$, the zeros of the Bessel function
$j_{m}^{(n)}\approx\pi n$. Thus the allowed values of the wave numbers are
$k_n^{\text{vac}}\approx \pi n/L$, similar to the 1D case. However, the
above--used simple equation for $j_{m}^{(n)}$  is valid only if $|m|$ is not
very large. In the case of $|m|\gg1$, the first zero of the Bessel function
$j_{m}^{(1)}\approx |m|$. Hence, in a finite system there appears a
restriction for the allowed number of modes, $|m|\leq L$, and the sum rule for
the 2D case takes the form:
\begin{equation} \label{eq:2Dsum-rules}
\sum_{k,m} \Bigl( \bullet \Bigr) \longrightarrow
\frac{L}{\pi}\int_0^{\infty}\mathrm{d}k \sum_{m=-kL}^{kL} \Bigl( \bullet
\Bigr).
\end{equation}
For magnon states in the presence of the soliton there appears a phase shift
$\eta_m$ due to the soliton--magnon scattering, therefore
\begin{equation*}
k_nL+\eta_m = k_n^{\text{vac}}L = j_{m}^{(n)}\qquad \Longrightarrow \quad
k_n-k_n^{\text{vac}} = -\frac{\eta}{L}.
\end{equation*}
The one-loop correction to the soliton energy reads
\begin{equation} \label{eq:1-loop-2}
\mathscr{E}^{\text{1--loop}} =  - \frac{\hslash D}{\pi} \!\!
\int_0^{\infty}\!\!\!\! k\,\eta(k) \mathrm{d}k,\quad \eta(k) = \!\!\!
\sum_{m=-kL}^{kL} \eta_m(k).\!\!\!
\end{equation}
Here $\eta(k)$ is a sum of the phase shifts of all partial waves with the
fixed radial wave number $k$. Note that our definition of $\eta(k)$, in
contrast to the case of the $\sigma$-model \cite{Walliser00}, contains
independent summations over positive and negative $m$, which reflects the
breaking of the symmetry $m\to-m$.

The short--wave length behavior of the phase shift is responsible for
ultraviolet (UV) singularities. Following Refs.~\onlinecite{Walliser00} in
order to avoid the UV singularity one needs to derive the short--wave length
asymptotics for the phase shift. There appears a problem to sum to infinity an
alternating series, which has no absolute convergence. Symmetric limits
regularize this summation similar to the exponential regularization for the
original AB--scattering \cite{Henneberger80}; different waves are taken into
account in the order in which the poles $k_m^{(p)}$ in the scattering
amplitude appear as $k$ increases \footnote{The pole of the scattering
amplitude corresponds to the condition when $\eta(k_m^{(p)})=\pm\pi/2$.
Asymptotically $k_m^{(p)}\sim |m|/R$ when $k>>1/R$, see details in
Ref.~\onlinecite{Ivanov99}.}. Finally, we found that $\eta(\infty) = -\pi$ for
the soliton with $q=1$\footnote{More general, $\eta(\infty) = -\pi|q|$ for
higher topological charges \cite{Sheka06d}.}. Note that this result can not be
obtained from the Born approximation because of the long--range nature of the
AB--scattering; namely, perturbative Born calculations resulted in different
conclusions that $\eta(\infty)=\pi$ (see Ref.~\onlinecite{Rodriguez89}) and
$\eta(\infty)=2\pi$ (see Ref.~\onlinecite{Walliser00}).

To check our analytical predictions we have calculated $\eta(k)$ by the
numerical integration of the Schr\"{o}dinger equation \eqref{eq:Schroedinger}
for $m\in[-100;100]$, which gives $\eta(\infty)=-\pi$ with the precision
$10^{-3}$.

\begin{figure}
\includegraphics[width=\columnwidth]{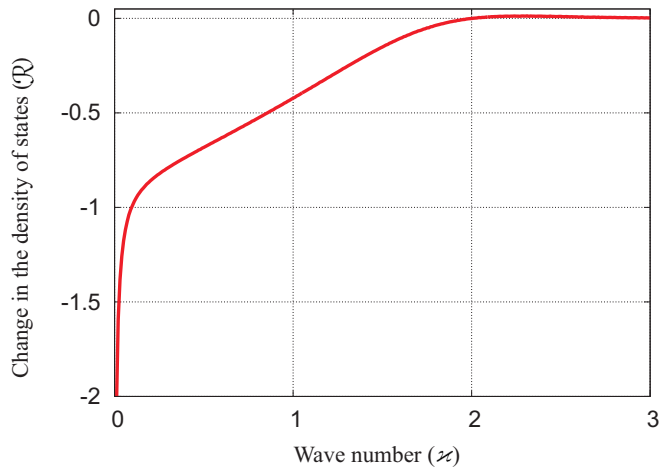}
\caption{The change $\mathcal{R}$ in the density of magnon states due to the
soliton \emph{vs} dimensionless wavenumber ($\varkappa=kR$), obtained
by numerical integration of the Schr\"{o}dinger equation \eqref{eq:Schroedinger}.}%
\label{fig:Rho} %
\end{figure}

Finally, the one-loop correction reads $\mathscr{E}^{\text{1--loop}}=
\mathscr{E}_{\text{CT}} + \mathscr{E}^{\text{Cas}}$. Here the counterterm
$\mathscr{E}_{\text{CT}} = -\left.\frac{\hslash D}{2\pi}\eta(k)
k^2\right|_0^\infty$ can be evaluated using a momentum cutoff $\Lambda$, which
results in $\mathscr{E}_{\text{CT}} = \frac12 \hslash D\Lambda^2$. Taking away
this UV term, e.g. by renormalizing the exchange constant \cite{Walliser00},
we ends with the finite Casimir energy:
\begin{equation} \label{eq:1-loop+Cas}
\mathscr{E}^{\text{Cas}}  =  \frac{\hslash D}{2R^2}\int\limits_0^{\infty}\!
\varkappa^2 \mathcal{R}(\varkappa)\mathrm{d}\varkappa, \
\mathcal{R}(\varkappa) =\frac{1}{\pi}\!\!\!\!\!\! \sum_{m=-\varkappa
L/R}^{\varkappa L/R}\!\!\! \frac{\mathrm{d} \eta_m(\varkappa)}{\mathrm{d}
\varkappa}.
\end{equation}
Here $\mathcal{R}(\varkappa)$ describes the change in the density of magnon
states due to the soliton. This expression can be analyzed analytically in
limiting cases. Using the asymptotical behavior for the phase shift for
different modes one can conclude that the maximum scattering in the long
wavelength limit corresponds to the mode with $m=0$, hence the density of
states has a singularity: $\mathcal{R}(\varkappa) \sim
-(2\varkappa)^{-1}\ln^2\varkappa\to -\infty$ when $\varkappa\to0$. In the
short wavelength limit all modes compensate each other and
$\mathcal{R}(\varkappa \to\infty)\to0$. In the intermediate range the change
in the density of states can be found numerically only, see
Fig.~\ref{fig:Rho}. One can conclude that for all $\varkappa$ the change in
the density of states takes only negative values and $\int_0^\infty
\mathcal{R}(\varkappa)\mathrm{d}\varkappa = -1$. Finally, the Casimir energy
is obtained as
\begin{equation} \label{eq:E-Cas4FM}
\begin{split}
&\frac{\mathscr{E}^{\text{Cas}}}{\mathscr{E}^{\text{BP}}} = - \frac{C}{8\pi S}
\left(\frac{a}{R}\right)^2,\quad C = \int\limits_0^\infty\!\! \varkappa^2
\left|\mathcal{R}(\varkappa)\right| \mathrm{d}\varkappa,
\end{split}
\end{equation}
where $C$ is a constant, which we calculated numerically to be $C \approx
0.38$. We introduced in  \eqref{eq:E-Cas4FM} the typical length scale
$a=\sqrt{\hbar D/JS}$, which is about a lattice constant. Note that this
parameter is absent for the static BP--soliton problem, but it naturally
breaks an initial scale invariance of the model in the dynamics, see
Eq.~\eqref{eq:LL}. The soliton energy is reduced when the soliton radius
decreases.

Let us discuss physical consequences of Eq.~\eqref{eq:E-Cas4FM}. First, the
soliton energy with account of the quantum correction does not have a minimum
as a function of the soliton radius. Usually this is interpreted as a soliton
instability in the context of the Hobart--Derrick theorem, see
Ref.~\onlinecite{Manton04}.  We will show here that this property leads to a
dissipation of the soliton energy  caused by magnon  radiation, common to that
for 3D Hopf solitons in isotropic FMs \cite{Dzyaloshinskii79}. As an important
contrast to $\sigma$--models, a model of the FM has an additional integral of
motion, the $z$--component of the total spin $S_z\approx 2\pi S(R/a)^2$
\cite{Kosevich90}. Even for the static limit $S_z$ takes a nonzero value; the
energy dissipation caused by the radiation of magnon pairs with wave vectors
$\vec{k}$ and $-\vec{k}$ will be accompanied by a decrease of the value of
$S_z$ by two. Therefore the soliton lifetime $\tau = S_z/(\mathrm{d}
S_z\!/\mathrm{d} t)$ can be sizeable, when $S_z$ is large ($S_z$ is the number
of bound magnons in the soliton).

The amplitude of the radiation process is $\varpi\sim JS (ak)^2$, see
\cite{Dzyaloshinskii79}. In accordance to Fermi's golden rule,
\begin{equation*} \label{eq:Fermi}
\frac{\mathrm{d}S_z}{\mathrm{d}t} = \sum_{k}2\pi \frac{|\varpi|^2}{\hslash}
\delta\left(\frac{\mathrm{d}E}{\mathrm{d} S_z} - \hslash \omega(k) \right).
\end{equation*}
Here $E$ is the total energy of the soliton with account of the Casimir
energy. Calculating $\mathrm{d}S_z/\mathrm{d}t$ in the continuum limit, one
can finally obtain the soliton lifetime
\begin{equation} \label{eq:lifetime}
\tau \sim \frac{4\pi\hslash}{J C^2} \left(\frac{R}{a}\right)^{10}.
\end{equation}
Note that the lifetime is much bigger than $\hslash/J$ for $R>a$.

To conclude, quantum effects decrease the energy of the BP soliton in
isotropic FMs, more strongly for small soliton radius. Nevertheless, the
original argumentation by \citet{Belavin75} about the breaking of the
long-range order of the system is still valid. It is based on the fact that in
isotropic FMs the energy is independent of the soliton radius $R$, and $R$ can
be comparable with the system size, $R\sim L$. However, in the case of large
radii the quantum correction is negligible, the energy has a finite limit for
$R\to\infty$, so the problem of long-range ordering has the classical form.
Another aspect of the problem is the fate of the BP soliton at small finite
$R$. As can be seen form Eq.~\eqref{eq:lifetime}, the lifetime is small for
small $R$. In some respects the situation is similar to the problem of the
black hole evaporation, i.e. the large radius soliton dissipation is very
slow, and can be neglected. At the same time the dissipation of small radius
solitons is very fast. When the soliton radius is small enough, the speed of
dissipation increases rapidly; in the final stage with $S_z\sim S$ the soliton
can collapse by a quantum jump, which is accompanied by a change of the
topological charge.

\begin{acknowledgments}
D.~D.~Sh. thanks the University of Bayreuth, where part of this work was
performed, for kind hospitality and acknowledges the support from the
Alexander von Humboldt Foundation. The work in Kiev is partly supported by the
grant INTAS Ref.05-1000008-8112.
\end{acknowledgments}

%\bibliography{soliton}

\begin{thebibliography}{23}
\expandafter\ifx\csname natexlab\endcsname\relax\def\natexlab#1{#1}\fi
\expandafter\ifx\csname bibnamefont\endcsname\relax
  \def\bibnamefont#1{#1}\fi
\expandafter\ifx\csname bibfnamefont\endcsname\relax
  \def\bibfnamefont#1{#1}\fi
\expandafter\ifx\csname citenamefont\endcsname\relax
  \def\citenamefont#1{#1}\fi
\expandafter\ifx\csname url\endcsname\relax
  \def\url#1{\texttt{#1}}\fi
\expandafter\ifx\csname urlprefix\endcsname\relax\def\urlprefix{URL }\fi
\providecommand{\bibinfo}[2]{#2}
\providecommand{\eprint}[2][]{\url{#2}}

\bibitem[{\citenamefont{Manton and Sutcliffe}(2004)}]{Manton04}
\bibinfo{author}{\bibfnamefont{N.}~\bibnamefont{Manton}} \bibnamefont{and}
  \bibinfo{author}{\bibfnamefont{P.}~\bibnamefont{Sutcliffe}},
  \emph{\bibinfo{title}{Topological solitons}}, Cambridge Monographs on
  Mathematical Physics (\bibinfo{publisher}{Cambridge University Press},
  \bibinfo{year}{2004}).

\bibitem[{\citenamefont{Kosevich et~al.}(1990)\citenamefont{Kosevich, Ivanov,
  and Kovalev}}]{Kosevich90}
\bibinfo{author}{\bibfnamefont{A.~M.} \bibnamefont{Kosevich}},
  \bibinfo{author}{\bibfnamefont{B.~A.} \bibnamefont{Ivanov}},
  \bibnamefont{and} \bibinfo{author}{\bibfnamefont{A.~S.}
  \bibnamefont{Kovalev}}, \bibinfo{journal}{Physics Reports}
  \textbf{\bibinfo{volume}{194}}, \bibinfo{pages}{117} (\bibinfo{year}{1990}),
  \urlprefix\url{http://www.sciencedirect.com/science/article/B6TVP-46SXP0P-19%
/2/d0a99bbbccf078602123db3e5d601202}.

\bibitem[{\citenamefont{Mikeska and Steiner}(1991)}]{Mikeska91}
\bibinfo{author}{\bibfnamefont{H.~J.} \bibnamefont{Mikeska}} \bibnamefont{and}
  \bibinfo{author}{\bibfnamefont{M.}~\bibnamefont{Steiner}},
  \bibinfo{journal}{Adv. Phys.} \textbf{\bibinfo{volume}{40}},
  \bibinfo{pages}{191} (\bibinfo{year}{1991}).

\bibitem[{\citenamefont{Bar'yakhtar and Ivanov}(1993)}]{Baryakhtar93}
\bibinfo{author}{\bibfnamefont{V.~G.} \bibnamefont{Bar'yakhtar}}
  \bibnamefont{and} \bibinfo{author}{\bibfnamefont{B.~A.}
  \bibnamefont{Ivanov}}, \bibinfo{journal}{Sov. Sci. Rev. Sec.A.}
  \textbf{\bibinfo{volume}{16}}, \bibinfo{pages}{3} (\bibinfo{year}{1993}).

\bibitem[{\citenamefont{Belavin and Polyakov}(1975)}]{Belavin75}
\bibinfo{author}{\bibfnamefont{A.~A.} \bibnamefont{Belavin}} \bibnamefont{and}
  \bibinfo{author}{\bibfnamefont{A.~M.} \bibnamefont{Polyakov}},
  \bibinfo{journal}{JETP Lett.} \textbf{\bibinfo{volume}{22}},
  \bibinfo{pages}{245} (\bibinfo{year}{1975}).

\bibitem[{\citenamefont{Abolfath}(1998)}]{Abolfath98}
\bibinfo{author}{\bibfnamefont{M.}~\bibnamefont{Abolfath}},
  \bibinfo{journal}{Phys. Rev. B} \textbf{\bibinfo{volume}{58}},
  \bibinfo{pages}{2013} (\bibinfo{year}{1998}),
  \urlprefix\url{http://link.aps.org/abstract/PRB/v58/p2013}.

\bibitem[{\citenamefont{Rodriguez}(1989)}]{Rodriguez89}
\bibinfo{author}{\bibfnamefont{J.~P.} \bibnamefont{Rodriguez}},
  \bibinfo{journal}{Phys. Rev. B} \textbf{\bibinfo{volume}{39}},
  \bibinfo{pages}{2906} (\bibinfo{year}{1989}),
  \urlprefix\url{http://link.aps.org/abstract/PRB/v39/p2906}.

\bibitem[{\citenamefont{Moussallam}(1991)}]{Moussallam91}
\bibinfo{author}{\bibfnamefont{B.}~\bibnamefont{Moussallam}},
  \bibinfo{journal}{Phys. Rev. B} \textbf{\bibinfo{volume}{43}},
  \bibinfo{pages}{3325} (\bibinfo{year}{1991}),
  \urlprefix\url{http://link.aps.org/abstract/PRB/v43/p3325}.

\bibitem[{\citenamefont{Walliser and Holzwarth}(2000)}]{Walliser00}
\bibinfo{author}{\bibfnamefont{H.}~\bibnamefont{Walliser}} \bibnamefont{and}
  \bibinfo{author}{\bibfnamefont{G.}~\bibnamefont{Holzwarth}},
  \bibinfo{journal}{Phys. Rev. B} \textbf{\bibinfo{volume}{61}},
  \bibinfo{pages}{2819} (\bibinfo{year}{2000}),
  \urlprefix\url{http://link.aps.org/abstract/PRB/v61/p2819}.

\bibitem[{\citenamefont{Dashen et~al.}(1974{\natexlab{a}})\citenamefont{Dashen,
  Hasslacher, and Neveu}}]{Dashen74I}
\bibinfo{author}{\bibfnamefont{R.~F.} \bibnamefont{Dashen}},
  \bibinfo{author}{\bibfnamefont{B.}~\bibnamefont{Hasslacher}},
  \bibnamefont{and} \bibinfo{author}{\bibfnamefont{A.}~\bibnamefont{Neveu}},
  \bibinfo{journal}{Phys. Rev. D} \textbf{\bibinfo{volume}{10}},
  \bibinfo{pages}{4114} (\bibinfo{year}{1974}{\natexlab{a}}),
  \urlprefix\url{http://link.aps.org/abstract/PRD/v10/p4114}.

\bibitem[{\citenamefont{Dashen et~al.}(1974{\natexlab{b}})\citenamefont{Dashen,
  Hasslacher, and Neveu}}]{Dashen74II}
\bibinfo{author}{\bibfnamefont{R.~F.} \bibnamefont{Dashen}},
  \bibinfo{author}{\bibfnamefont{B.}~\bibnamefont{Hasslacher}},
  \bibnamefont{and} \bibinfo{author}{\bibfnamefont{A.}~\bibnamefont{Neveu}},
  \bibinfo{journal}{Phys. Rev. D} \textbf{\bibinfo{volume}{10}},
  \bibinfo{pages}{4130} (\bibinfo{year}{1974}{\natexlab{b}}),
  \urlprefix\url{http://link.aps.org/abstract/PRD/v10/p4130}.

\bibitem[{\citenamefont{Rajaraman}(1982)}]{Rajaraman82}
\bibinfo{author}{\bibfnamefont{R.}~\bibnamefont{Rajaraman}},
  \emph{\bibinfo{title}{Solitons and {I}nstanton}}
  (\bibinfo{publisher}{North--Holland}, \bibinfo{address}{Amsterdam},
  \bibinfo{year}{1982}).

\bibitem[{\citenamefont{Ivanov et~al.}(1999)\citenamefont{Ivanov, Murav'ev, and
  Sheka}}]{Ivanov99}
\bibinfo{author}{\bibfnamefont{B.~A.} \bibnamefont{Ivanov}},
  \bibinfo{author}{\bibfnamefont{V.~M.} \bibnamefont{Murav'ev}},
  \bibnamefont{and} \bibinfo{author}{\bibfnamefont{D.~D.} \bibnamefont{Sheka}},
  \bibinfo{journal}{JETP} \textbf{\bibinfo{volume}{89}}, \bibinfo{pages}{583}
  (\bibinfo{year}{1999}),
  \urlprefix\url{http://link.aip.org/link/?JET/89/583/1}.

\bibitem[{\citenamefont{Sheka and Mertens}(2006)}]{Sheka06d}
\bibinfo{author}{\bibfnamefont{D.~D.} \bibnamefont{Sheka}} \bibnamefont{and}
  \bibinfo{author}{\bibfnamefont{F.~G.} \bibnamefont{Mertens}},
  \bibinfo{journal}{Phys. Rev. A} \textbf{\bibinfo{volume}{74}},
  \bibinfo{eid}{052703} (pages~\bibinfo{numpages}{5}) (\bibinfo{year}{2006}),
  \urlprefix\url{http://link.aps.org/abstract/PRA/v74/e052703}.

\bibitem[{\citenamefont{Sheka et~al.}(2003)\citenamefont{Sheka, Ivanov, and
  Mertens}}]{Sheka03}
\bibinfo{author}{\bibfnamefont{D.}~\bibnamefont{Sheka}},
  \bibinfo{author}{\bibfnamefont{B.}~\bibnamefont{Ivanov}}, \bibnamefont{and}
  \bibinfo{author}{\bibfnamefont{F.~G.} \bibnamefont{Mertens}},
  \bibinfo{journal}{Phys. Rev. A} \textbf{\bibinfo{volume}{68}},
  \bibinfo{eid}{012707} (pages~\bibinfo{numpages}{5}) (\bibinfo{year}{2003}),
  \urlprefix\url{http://link.aps.org/abstract/PRA/v68/e012707}.

\bibitem[{\citenamefont{Henneberger}(1980)}]{Henneberger80}
\bibinfo{author}{\bibfnamefont{W.~C.} \bibnamefont{Henneberger}},
  \bibinfo{journal}{Phys. Rev. A} \textbf{\bibinfo{volume}{22}},
  \bibinfo{pages}{1383} (\bibinfo{year}{1980}).

\bibitem[{\citenamefont{Feinberg}(1963)}]{Feinberg63}
\bibinfo{author}{\bibfnamefont{E.}~\bibnamefont{Feinberg}},
  \bibinfo{journal}{Sov. Phys. Usp.} \textbf{\bibinfo{volume}{5}},
  \bibinfo{pages}{753} (\bibinfo{year}{1963}),
  \urlprefix\url{http://www.ufn.ru/archive/russian/abstracts/abst3970.html}.

\bibitem[{\citenamefont{Ruijsenaars}(1983)}]{Ruijsenaars83}
\bibinfo{author}{\bibfnamefont{S.~N.~M.} \bibnamefont{Ruijsenaars}},
  \bibinfo{journal}{Annals of Physics} \textbf{\bibinfo{volume}{146}},
  \bibinfo{pages}{1} (\bibinfo{year}{1983}),
  \urlprefix\url{http://www.sciencedirect.com/science/article/B6WB1-4DDR3KP-NJ%
/2/d8f56ee8b6ab36e8635946f3f839dfbf}.

\bibitem[{\citenamefont{Sheka et~al.}(2001)\citenamefont{Sheka, Ivanov, and
  Mertens}}]{Sheka01}
\bibinfo{author}{\bibfnamefont{D.~D.} \bibnamefont{Sheka}},
  \bibinfo{author}{\bibfnamefont{B.~A.} \bibnamefont{Ivanov}},
  \bibnamefont{and} \bibinfo{author}{\bibfnamefont{F.~G.}
  \bibnamefont{Mertens}}, \bibinfo{journal}{Phys. Rev. B}
  \textbf{\bibinfo{volume}{64}}, \bibinfo{pages}{024432}
  (\bibinfo{year}{2001}),
  \urlprefix\url{http://link.aps.org/abstract/PRB/v64/e024432}.

\bibitem[{\citenamefont{Sheka et~al.}(2004)\citenamefont{Sheka, Yastremsky,
  Ivanov, Wysin, and Mertens}}]{Sheka04}
\bibinfo{author}{\bibfnamefont{D.~D.} \bibnamefont{Sheka}},
  \bibinfo{author}{\bibfnamefont{I.~A.} \bibnamefont{Yastremsky}},
  \bibinfo{author}{\bibfnamefont{B.~A.} \bibnamefont{Ivanov}},
  \bibinfo{author}{\bibfnamefont{G.~M.} \bibnamefont{Wysin}}, \bibnamefont{and}
  \bibinfo{author}{\bibfnamefont{F.~G.} \bibnamefont{Mertens}},
  \bibinfo{journal}{Phys. Rev. B} \textbf{\bibinfo{volume}{69}},
  \bibinfo{eid}{054429} (pages~\bibinfo{numpages}{13}) (\bibinfo{year}{2004}),
  \urlprefix\url{http://link.aps.org/abstract/PRB/v69/e054429}.

\bibitem[{\citenamefont{Ivanov and Sheka}(2005)}]{Ivanov05b}
\bibinfo{author}{\bibfnamefont{B.~A.} \bibnamefont{Ivanov}} \bibnamefont{and}
  \bibinfo{author}{\bibfnamefont{D.~D.} \bibnamefont{Sheka}},
  \bibinfo{journal}{JETP Lett.} \textbf{\bibinfo{volume}{82}},
  \bibinfo{pages}{436} (\bibinfo{year}{2005}),
  \urlprefix\url{http://link.aip.org/link/?JTP/82/436/1}.

\bibitem[{\citenamefont{Ivanov}(1995)}]{Ivanov95g}
\bibinfo{author}{\bibfnamefont{B.~A.} \bibnamefont{Ivanov}},
  \bibinfo{journal}{JETP Lett.} \textbf{\bibinfo{volume}{61}},
  \bibinfo{pages}{917} (\bibinfo{year}{1995}),
  \urlprefix\url{http://www.jetpletters.ac.ru/ps/936/article_14292.shtml}.

\bibitem[{\citenamefont{Dzyaloshinskii and Ivanov}(1979)}]{Dzyaloshinskii79}
\bibinfo{author}{\bibfnamefont{I.~E.} \bibnamefont{Dzyaloshinskii}}
  \bibnamefont{and} \bibinfo{author}{\bibfnamefont{B.~A.}
  \bibnamefont{Ivanov}}, \bibinfo{journal}{JETP Lett}
  \textbf{\bibinfo{volume}{29}}, \bibinfo{pages}{592} (\bibinfo{year}{1979}).

\end{thebibliography}

\end{document}